\documentclass[a4paper,twocolumn]{esapub} 
\usepackage{natbib,graphicx}
\usepackage{amssymb,multirow}
\newcommand{\al}{$^{26}$Al~}
\def\msun{${\rm M}_{\odot}$}
\def\kms{${\rm km}.{\rm s}^{-1}$}

\title{Nucleosynthesis of \al in rotating Wolf-Rayet stars}
\author{A. Palacios}
\affil{Institut d'Astronomie et d'Astrophysique - U.L.B. - CP-226, Bd du
              Triomphe, B-1050 Brussels (Belgium), {\sf palacios@astro.ulb.ac.be}}
\author{G. Meynet}
\affil{Observatoire de Gen\`eve, CH-1290 Sauverny (Switzerland), {\sf Georges.Meynet@obs.unige.ch}}
\author{C. Vuissoz}
\affil{ Laboratoire d'Astrophysique de l'EPFL - Observatoire - CH-1290 Chavannes-des-Bois (Switzerland), {\sf Christel.Vuissoz@obs.unige.ch}}

\begin{document}


\maketitle

\begin{abstract}
The \al radionuclide can be detected through its decay emission line at
1.809 MeV, as was first observed by Mahoney et al. (1982). Since
then, COMPTEL on board of the CGRO satellite, performed a sky survey in
this energy range, and provided maps of the \al distribution in the Galaxy.
These results revealed that the main contributors to the synthesis
of \al are most likely the massive stars, which contribute through their winds (Wolf-Rayet
stars) and through their supernova explosion.\\ 
Comparison between these observations (in particular observations in localized
regions such as the Vela region and the Cygnus region) and the models
available at that moment, showed however the need for improvements from
both theoretical and observational points of view, in order to improve
our understanding of the \al galactic distribution as well as that of its synthesis.\\
With the launch of the INTEGRAL satellite in October 2002, the
observational part will hopefully be improved, and the construction of
better resolution maps at 1.809 MeV is one of the main aims of the
mission. From a theoretical point of view, we need the most up-to-date
predictions in order to be able to interpret the forthcoming data.\\
In this paper, we address this latter part, and present new results for \al
production by rotating Wolf-Rayet stars and their contribution to the total
amount observed in the Galaxy. 
\end{abstract}

\section{\al sources}

\al is a radioactive nuclide that can be produced by
hydrostatic nucleosynthesis in H burning regions, by explosive
nucleosynthesis and by spallation. Its half--life time in its ground state
is of $7.2 \times 10^5$ yr, and it is thus a good tracer of recent nucleosynthetic
events in the Galaxy.\\
The maps obtained with COMPTEL (Diehl et al. 1995, Kn\"odlseder et
al. 1999, Pl\"uschke et al. 2001) allowed to pin down the main contributors to
the diffuse emission observed mainly in the galactic plane : massive
stars (Prantzos \& Diehl 1996, Kn\"odlseder 1999). These objects end their lives with a supernova explosion, during
which \al can be synthesized and will eventually be expelled (Heger et al. 2003). The more
massive (and shorter lived) ones will also contribute during quiescent
evolutionary phases through their winds.

\section{Wolf-Rayet stars}

Wolf-Rayet (hereafter WR) stars are the bare cores of massive O--type stars
that lost their hydrogen rich envelope due to very strong stellar
winds or by mass transfer in a close
binary system. They exhibit at their
surface the products of the internal nuclesoynthesis accompanying first
the central H--burning (WN phase) and then central He--burning phase (WC and WO phases).
At solar metallicity, all stars more massive than about 30 \msun~go
through a WR phase, which typically lasts for a few $10^5$ years in the
case of non-rotating objects (Meynet \& Maeder 2003; 2004), and
massively contribute to the chemical enrichment of the interstellar
medium (hereafter ISM).\\ 
\al is synthesized during the central H--burning phase in these
stars, and is eventually ejected into the ISM as the star is peeled
off by strong winds during the WR phase.
The amount of \al ejected in the ISM through winds is very sensitive
to mass loss rates, metallicity, rotation and initial mass. Meynet et
al. (1997) presented a first study of the effects of initial mass and
metallicity, as well as the effects of different possible
prescriptions for convection. At that time, they could already show
that the amount of \al ejected increases with metallicity and initial
mass. In the grid of models used by Meynet et al. (1997), artificially
twice enhanced mass loss rates were used, which appeared at that time to better fit the observed
properties of the WR stars in a series of environments (Maeder \& Meynet 1994).  

In the following, we present the results obtained for a new grid of models
including rotation and updated mass loss prescriptions for the different
evolutionary phases that massive stars are likely to go through. Part
of these results were discussed in Vuissoz et al. (2004). 

\section{Rotation induced mixing}

Massive stars are known to have high equatorial velocities ($<$v$>$ = 200
-- 250 \kms)
while on main sequence.
Such a rotation affects the evolution of stars in several ways. It
first has a direct incidence on its structure through the centrifugal acceleration.  
It also affects the mass loss in terms of geometry and
quantity. Finally it triggers instabilities in the radiative interior
of the star, which allow transport of angular momentum and chemical
species.

Transport of angular momentum and chemical species is ensured by the
meridional circulation, which settles in to balance the thermal
desequilibrium induced by the departure from spherical symetry in rotating
objects, and by the shear instability associated with the turbulent
layers in regions with high angular velocity gradients (Zahn
1992; Maeder \& Zahn 1998).

We chose an initial velocity $\upsilon =$ 300 \kms~ for all masses
and metallicities, taking into account that this value is compatible
with the mean averaged observed velocity of O-type stars at solar
metallicity. Considering this, the rotating models used here well
reproduce a series of observed features for O-type and WR stars, that
non--rotating models fail to fit (Meynet \& Maeder 2004).

\section{Results for \al}

\begin{figure}
\centering
\includegraphics[height=8cm,width=8cm]{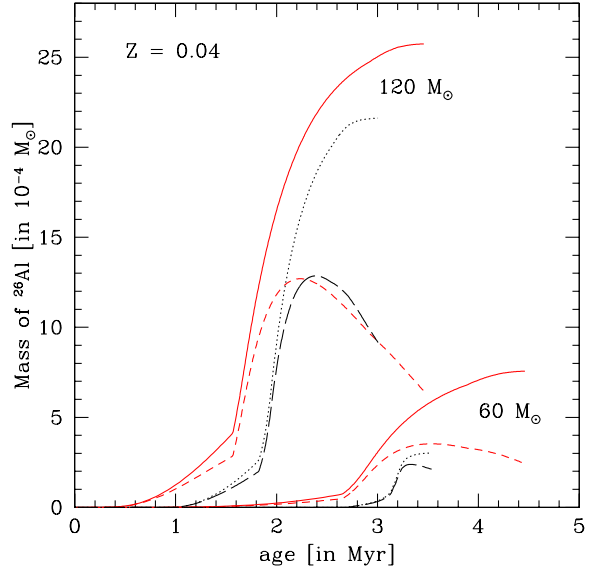}
\caption{Evolution as a function of time of the cumulated mass of \al ejected by WR stellar winds at twice
the solar metallicity. Continuous line show the case of rotating models,
dotted lines the case of non--rotating models. The short-dashed lines and
the  long--dashed lines
show the evolution as a function of time of the cumulated mass of ``live'' \al, {\it i.e.}
of still non--decayed \al, for rotating and non--rotating cases respectively..}
\end{figure}

\begin{figure}
\centering
\includegraphics[height=8cm,width=8cm]{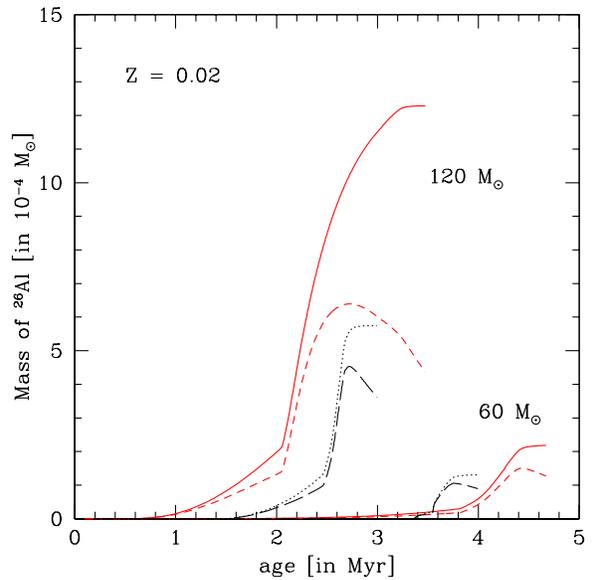}
\caption{Same as Fig. 1 for solar metallicity models.}
\end{figure}

The effects of rotation on \al production in WR stars are consequences of
both the effects on the mass loss and of the rotation induced mixing.
The main results for what concerns \al are the following :
\begin{itemize}
\item At given metallicity, rotation allows stars with lower initial
  masses to enter the WR  phase. When we take stellar rotation into account
  (which is an observational fact), the number of stars contributing to the \al enrichment of the
  ISM through stellar winds is thus increased.  The total contribution of
  WR stars to the galactic content of \al is larger when using rotating
  models than when using non--rotating ones, as can be seen in Fig. 3, and
  as will be discussed in the next section. 
\item At given mass and metallicity, rotation results in an entrance into the
  WR phase at an earlier evolutionary stage (not necessarily at an earlier time).
  For instance the solar metallicity 60 M$_\odot$ star model enters the WR phase
  during the core He--burning phase when it is non--rotating and during the end of
  the core H--burning phase when it is rotating. In this last case the duration 
  of the WR phase is longer so that the cumulative
  time interval during which \al is expelled is larger for rotating
  stars. This leads to an enhancement of the \al yields. This is well
  illustrated in Figs. 1 and 2 where the evolution of the cumulated
  mass of \al is shown as a function of time for both rotating
  and non--rotating models.
\item The higher the metallicity, the larger the ejected mass
of \al (see Fig.~1 and Fig.~2).\\ 
This is due to the metallicity dependence of mass loss
  rates and to the higher $^{25}{\rm Mg}$ mass fraction available for \al
  synthesis at higher
  metallicity. 
\item  Rotation allows surface enrichment of \al prior to the WR
  phase. This is a direct consequence of the combined effects of meridional
  circulation and turbulence which allow \al to diffuse outwards from the
  convective core, where it is produced by proton capture on $^{25}{\rm
  Mg}$, up to the radiative envelope,
  and to appear earlier at the surface of the star. 
\end{itemize}
\begin{figure*}[tH]
\begin{center}
\includegraphics[height=9cm,width=9cm]{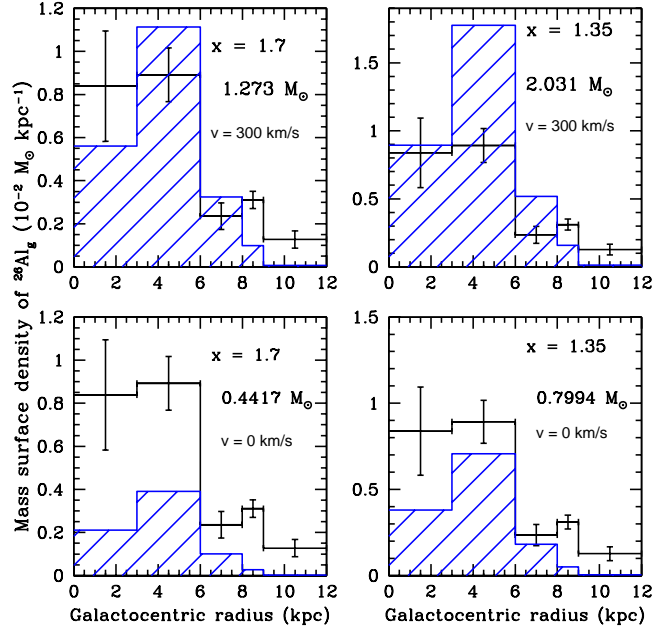}
\caption{Theoretical profiles of the radial surface density of \al in the
  Galaxy (hatched histograms) compared with the observational profile obtained
  after COMPTEL data analysis (Kn\"odlseder 1997) (black
  line). The total \al mass ejected by WR winds in the whole Galaxy during the last million years
  (in \msun ) is indicated on
  each plot. \underline {Upper panels}:
  Profiles obtained for rotating models $\upsilon_{\rm ini}$ = 300 \kms,
  for a ``Scalo'' IMF slope $x = 1.7$ (on the left), and a
  ``Salpeter'' IMF slope $x = 1.35$ (on the right).
  \underline {Lower panels}: Same as upper panels, but using non--rotating models.}
\end{center}
\end{figure*}

\section{Contribution of WR stars to the \al galactic content}

One can evaluate the contribution of WR stars to the total galactic mass of
\al in the following way:
\footnotesize
 \begin{eqnarray}
 M_{26}^{gal} & = & \tau_{26} . \dot{M}^{gal}_{26} \approx  \dot{M}^{gal}_{26} \nonumber\\ 
 & = & \int^R_0 2 \pi r dr \sigma (r)
 \int^{120}_{M_{\rm LWR}(r,Z)} \Phi(M) Y_{26}^W (M_i, r, Z) dM \nonumber\\
 & &  ~~~~{\rm if}~  \tau_{26} = 1.44~{\rm Myr} \approx  1~{\rm Myr},  
 \end{eqnarray}
\normalsize
R is the galactic radius taken equal to 15 kpc.\\ $\sigma(r)$ is the
surface density distribution of atomic and molecular hydrogen according to Scoville \& Sanders
(1987). This quantity is used to mimic the present star formation rate in the
Galaxy, provided that the molecular clouds distribution is considered a good tracer of
the star forming regions.\\ $M_{\rm LWR}$ is the minimum mass for a star to enter the WR phase at a
given metallicity. Its dependence on initial velocity and initial
metallicity is explicitely taken into account.\\ $\Phi(M) \propto {\rm m}^{-(1+x)}$ is the Initial Mass
Function, $x$ defining the slope of the IMF. Here, two different
prescriptions were used, in order to show the impact of this physical
ingredient on the global production of \al by WR stars.\\ $Y_{26}^W$ ($M_i$, r, Z) is the
yield of \al (excluding the supernova contribution) for a given mass $M_i$
at a given metallicity Z.

The expression in Eqn. 1 is normalized so as to account for a rate of 3 SNae per
century in the Galaxy (e.g. Capellaro et al. 1999).\\ Furthermore, we considered a metallicity gradient along the galactic radius
such as Z = 0.04 for r $\leq$ 4 kpc, and d log(Z)/dr = -0.07
dex.kpc$^{-1}$ beyond, with the condition that Z = 0.02 at r = 8.5 kpc
(solar neighbourhood).

In Fig. 3, we can see that the radial surface density profile of \al has
a similar shape whether we use rotating or non-rotating models. In
particular, in both cases there is a peak between 4 kpc and 6 kpc associated
with the existence of the molecular clouds ring at 5 kpc, which is
accounted for in the expression of the surface density distribution $\sigma(r)$.
\begin{itemize}
  \item The rotating models lead to a larger contribution of WR stars to the \al
  surface density profile in the inner 8 kpc ring. There, Z $\gtrsim {\rm
  Z}_{\odot}$, and as seen in Fig. 1 and Fig. 2, the combined effects of
  metallicity and rotation lead to a substantial enhancement of the yields.
   \item The total mass of \al originating from WR stars is increased by a factor 2.5 to 2.9
  when rotation is taken into account. The values obtained are also quite
  sensitive to the IMF slope, the Salpeter IMF ($x = 1.35$) favouring a larger
  contribution, even though the Scalo IMF ($x = 1.7$) is considered better suited for
  massive stars.
  \item For an IMF slope of 1.7, we obtain a total ejected mass of \al by WR
  stars of 1.27 \msun~ in case of rotating models, which represents  42\%
  to 63 \% of the total observed mass, that is estimated to range between 2
  and 3 \msun. 
\end{itemize}
  According to the recent observations by RHESSI (a $\gamma$-ray
  observatory for the Sun which was able to measure the diffuse emission of
  both \al and $^{60}{\rm Fe}$), the new estimate derived for the line ratio
  $^{60}{\rm Fe}/^{26}{\rm Al} \sim  0.1$ (see Smith's contribution in
  these proceedings) is consistent with the value
  predicted by Timmes et al. (1995), from computations including the sole
  contribution of type II supernovae (SNII). This would designate SNII as
  the dominant contributors to the \al galactic content. However, the most recent computations
  of nucleosynthesis from SNII indicate that the line ratio $^{60}{\rm
  Fe}/^{26}{\rm Al}$ should be larger than about 0.4. If this is the case,
  as pointed out by  Prantzos (2004; see also his contribution in these
  proceedings), \al should, at least up to a 50 \% level, originate from
  sources that produce \al but do not produce $^{60}{\rm Fe}$, as is the
  case for the WR stars. Our results show that rotating stellar models of
  WR stars could actually be responsible for such an amount of \al.
  Let us stress however, that this result has to be taken with caution,
  considering the 2.6 sigma level of the RHESSI detection of $^{60}{\rm
  Fe}$ and the large uncertainties existing on the physical parameters
  entering the derivation of such a quantity (see also Palacios et al., 2004). 

\section{Conclusions}

We have presented new results concerning \al production by WR stars from a
new grid of models including rotation and updated physics, in particular
the most recent prescriptions for mass loss rates.\\
Taking rotation into account globally leads to an enhancement of wind
ejected mass of \al by very massive stars.\\
Convolved with appropriate IMF and star formation rate indicator, these
yields lead to a total mass of about 1.3 \msun~of \al originating from WR
stellar winds. This value is in agreement with the conclusions drawn by
Prantzos (2004) from recent measurements of the line flux ratio $^{60}{\rm Fe}/^{26}{\rm Al}$.


\begin{thebibliography}{}

\bibitem[Capellaro 1999]{C99}
Capellaro, E., Evans, R., Turatto, M., 1999, A\&A, 351, 459

\bibitem[Diehl et al. 1995]{D95}
Diehl, R., Dupraz, C., Bennett, K., et al., 1995, A\&A, 298, 445

\bibitem[Heger et al. 2003]{H03}
Heger, A., Fryer, C. L., Woosley, S. E., Langer, N., Hartmann, D. H., 2003,
ApJ, 591, 288

\bibitem[Kn\"odlseder 1999a]{K99a}
Kn\"odlseder, J., 1999, ApJ, 510, 915

\bibitem[Kn\"odlseder 1999b]{K99b}
Kn\"odlseder, J., Dixon, D., Bennett, K., et al., 1999, A\&A,
345, 813

\bibitem[Kn\"odlseder 1997]{K97}
Kn\"odlseder, J., 1997, {\it The origin of \al in the Galaxy}, PhD thesis,
Paul Sabatier University, Toulouse (France)

\bibitem[Maeder \& Meynet 2000]{MM00}
Maeder A., Meynet G., 2000, A\&A, 361, 159

\bibitem[Maeder \& Meynet 1994]{MM94}
Maeder A., Meynet G., 1994, A\&A, 287, 803

\bibitem[Mahoney et al. 1982]{Ma1982}
Mahoney, W. A., Ling, J. C., Jacobson, A. S., Lingenfelter, R. E., 1982,
ApJ, 262, 742

\bibitem[Meynet et al. 1997]{M97}
Meynet, G., Arnould, M., Prantzos, N., Paulus, G., 1997, A\&A, 320, 460

\bibitem[Meynet \& Maeder 2004]{MM04}
Meynet, G., Maeder, A., 2004, submitted to A\&A

\bibitem[Meynet \& Maeder 2003]{MM03}
Meynet, G., Maeder, A., 2003, A\&A, 404, 975

\bibitem[Palacios et al. 2004]{P04}
Palacios, A., Meynet, G., Vuissoz, C., Kn\"odlseder, J., Cervi\~no, M.,
Schaerer, D., Mowlavi, N., 2004, A\&A , in preparation

\bibitem[Pl\"uschke et al. 2001]{P01}
Pl\"uschke, S., Diehl, R., Sch\"onfelder, et al., {\it The COMPTEL 1.809 MeV survey}, 2001, Proceedings of
the Fourth INTEGRAL Workshop, ESA SP-459, Noordwijk, 55 

\bibitem[Prantzos 2004]{P04}
Prantzos, N., 2004, A\&A, 420, 1033

\bibitem[Prantzos \& Diehl 1996]{PD96}
Prantzos, N., Diehl, R., 1996, Phys. Rep., 267, 1

\bibitem[Scoville \& Sanders 1987]{SS87}
Scoville, N. Z., Sanders, D. B.. 1987, ``{\it H2 in the Galaxy}'', in Interstellar Processes,
ed. D. Hollenbach \& H. Thronson (Dordrecht: Reidel), 21

\bibitem[Smith 2004]{S04}
Smith, D. M., 2004, this volume

\bibitem[Vuissoz et al. 2004]{V04}
Vuissoz, C.,  Meynet, G.,  Kn\"odlseder, J.,  Cervi\~no, M.,  Schaerer, D.,
Palacios, A.,  Mowlavi, N., 2004, New Astronomy Reviews, 48, 7
\end{thebibliography}
\end{document}